\documentclass[10pt,a4paper]{article}
\usepackage[latin1]{inputenc}
\usepackage{amsmath}
\usepackage{amsfonts}
\usepackage{amssymb}

\begin{document}



\title{The Hawking effect for massive particles}

\author{Bernard R. Durney}
\date{2377 Route de Carc\'{e}s, F-83510 Lorgues, France durney@physics.arizona.edu} 

\maketitle

\smallskip
\smallskip\textbf{Abstract}
This paper describes a particularly transparent derivation of the Hawking 
effect for massive particles in black holes. The calculations are 
performed with the help of Painlev\'{e}-Gullstrand's coordinates which are 
associated with a radially free-falling observer that starts at rest from 
infinity. It is shown that if the energy per unit rest mass, $e,$ is 
assumed to be related to the the Killing constant, $k,$ by $k^{2}=2e-1,$ 
then $e,$ must be greater than $\frac{1}{2}$. For particles that are 
confined below the event horizon (EH), $k$ is negative. In the quantum 
creation of particle pairs at the EH with $k=1$, the time component 
of the particle's four velocity that lies below the EH is compatible only 
with the time component of an outgoing particle above the EH, i.e, 
the outside particle cannot fall back on the black hole. Energy conservation 
requires that the particles inside, and outside the EH have the same value 
of $e$, and be created at equal distances from the EH, 
$(1-r_{in}=r_{out}-1).$ Global energy conservations forces then the mass 
of the particle below the EH to be negative, and equal to minus the mass 
the particle above the EH, i.e, the black hole looses energy as a 
consequence of pair production.

\smallskip

\smallskip \textbf{1. The Metric}

A transparent derivation of Painlev\'{e}-Gullstrand's coordinates (used in
this paper) is the following: Consider a radially, free-falling observer
that starts at rest from infinity. For this observer, the equations for $r$
as a function of the proper time $\tau $, and the equation for the
Schwarzschild time, $t$, as a function of $r$, are respectively (cf. Hartle
Eqs. (9.38), (9.40)),

$$
r\left( \tau \right) =\left( \frac{3}{2}\right) ^{\frac{2}{3}}\left(
2M\right) ^{\frac{1}{3}}\left( \tau _{\ast }-\tau \right) ^{\frac{2}{3}}
\eqno{(1.a)}
$$

$$
t(r)=t_{\ast }+2M\left\{ -\left( \frac{2}{3}\right) \left( \frac{r}{2M}%
\right) ^{\frac{3}{2}}-2\left( \frac{r}{2M}\right) ^{\frac{1}{2}}+\log |%
\frac{N}{D}|\right\} .
\eqno{(1.b)}
$$

$$
N=\left( \frac{r}{2M}\right) ^{\frac{1}{2}}+1,{ \ }D=\left( \frac{r}{2M}%
\right) ^{\frac{1}{2}}-1,
\eqno{(1.c)}
$$
where $\tau _{\ast }$ and $t_{\ast }$ are two constants. The equations
are written in geometrized units. It follows from Eq. (1.a) that the second
term in Eq.(1.b) is equal to $\tau -\tau _{\ast }$. This relation suggests
the following transformation of coordinates,

$$
t=\tau +2M\left\{ -2\left( \frac{r}{2M}\right) ^{\frac{1}{2}}+\log |\frac{N}{%
D}|\right\} 
\eqno{(2)}
$$
the constants in Eqs. (1.a) and (1.b) have been ignored because only $dt$ is
important. From Eq. (2) we obtain,

$$
dt=d\tau -dr\left( \frac{2M}{r}\right) ^{\frac{1}{2}}\left( 1-\frac{2M}{r}%
\right) ^{-1}. 
\eqno{(3.a)}
$$
which can also be derived directly from the equations,

$$
\frac{dr}{d\tau }=-w,{ \ }\frac{dt}{dr}=-\frac{1}{w\left(
1-w^{2}\right) },{ \ }w=\left( \frac{2M}{r}\right) ^{\frac{1}{2}}, 
\eqno{(3.b)}
$$
valid for a radial plunge with no kinetic energy at infinity. The new metric
can be written,

$$
ds^{2}=-\left( 1-\frac{2M}{r}\right) d\tau ^{2}+2\left( \frac{2M}{r}\right)
^{\frac{1}{2}}drd\tau +dr^{2} +r^2(d\theta^2 + \sin^2\theta d\phi^2)
\eqno{(4)}
$$

The radially ingoing and outgoing light rays are found to be ($d\tau /dr;$ 
it follows from the geodesic equations that $r$ is an affine parameter),

$$
in=-\frac{1}{\left( 1+w\right) },{ \ }out=\frac{1}{\left( 1-w\right) },%
{ \ }w=\left( \frac{2M}{r}\right) ^{\frac{1}{2}}. 
\eqno{(5)}
$$

The condition $ds^{2}=-d\tau^{2}$ requires that,

$$
\left( \frac{2M}{r}\right) d\tau ^{2}+2\left( \frac{2M}{r}\right)^{\frac{1}{2}}
 drd\tau +dr^{2}=0  
\eqno{(6)}
$$
the value of $dr/d\tau \left( =-w\right) $ obtained from Eq. (6) agrees with
Eq. (1.a).

\smallskip

\smallskip

\textbf{2. Particle Orbits}

\smallskip

\textit{Expression for a particle's energy per unit rest mass.}

\smallskip

We introduce the following change of notation: the time coordinate, $\tau $,
in Eq. (4) for the metric will be designated hereafter by $t$, whereas $\tau 
$\ will be a particle's proper time. For radial motions the equations for $%
u^{r}=dr/d\tau $ and $u^{t}=dt/d\tau $ are,

$$
{\bf{u}}\cdot {\bf{kv}}=\left( \frac{2M}{r}-1\right)
u^{t}+wu^{r}=-k,{ \ }\left( u^{r}\right) ^{2}=k^{2}+\frac{2M}{r}-1,%
{ \ }w=\left( \frac{2M}{r}\right) ^{\frac{1}{2}} 
\eqno{(7)}
$$
\smallskip where \textbf{u} = $(u^{t},u^{r})$, \textbf{kv} is the Killing
vector, namely (1, 0), and $k$ is a constant (the Killing constant). For
metrics that are independent of $t$, an integral of the geodesic equation
exists, that is given by the scalar product of (1, 0) with \textbf{u}, which
is Eq. (7.1), namely the first equation in (7). Equation (7.2) follows then
from the normalization condition for \textbf{u}, i.e., \textbf{u}$^{2}=-1$.
In the derivation of this equation, terms in the product $ku^{r}$ appear,
but they cancel. In Eq. (7.1), $w$ is dimensionless and it is clear that one
can assign to $k$ the dimension of a velocity (if dimensions are introduced
the first term would be multiplied by $c$.) We assume that $k$ and $e$, the
energy per unit rest mass, are related by, $k^{2}=2e-1$, and find that Eq.
(7.2), i.e., the second equation in (7), can be written,

$$
e=1+\frac{\left( u^{r}\right) ^{2}}{2}-\frac{M}{r}.
\eqno{(8)}
$$
\smallskip \qquad Little need is there to praise Eq. (8)! Because $k$ is
constant it follows from $k^{2}=2e-1$, that $e$ is also a constant, and if $%
k=1$, also $e=1$.

Equation (8) is an expression for the conservation of the particle's energy
as it moves along a time-independent space-time geometry.

\smallskip

\textit{Particle orbits for some important values of k.}

\smallskip

In this section we study particle orbits only for values of $k=0,1$, because
then (unlike other cases as, e.g., for $1 > k > 0$, in particular) solutions 
can easily be found. If $k=0$, and for radial motions, the equation for 
\textbf{u} is found to be,

$$
{\bf{u}}=\left\{ w\left( w^{2}-1\right) ^{-\frac{1}{2}},{ }%
-\left( w^{2}-1\right) ^{\frac{1}{2}},{ }0,{ }0\right\} ,{ }%
k=0 
\eqno{(9)}
$$
We expect that for physically meaningful solutions the particle's proper
time increases with the coordinate time, and have therefore discarded the
solution with $dt/d\tau =-w\left( w^{2}-1\right) ^{-\frac{1}{2}}.$ Equation
(9) is valid only below the event horizon and it follows from Eqs. (7.2) and
(8) that the value of $-u^{r}$ given by Eq. (9) is the minimum fall velocity,
and furthermore  that $e=\frac{1}{2}$. Therefore inside the event horizon, 
$e$ must be larger than $\frac{1}{2}$. Below the event horizon, the potential 
energy decreases, but this decrease is compensated by the increase in the 
minimum allowed value for $u^{r}$.

We proceed now to calculate the orbits for $|k|=1$. The value $k=1$,
corresponds to the orbit of a free-falling observer that starts at rest from
infinity. The value of $e$ for all the solutions with $|k|=1$ is unity. It
will be shown that $k=-1$, defines the orbit of a particle, with the same
energy per unit mass as the free falling observer, but confined below the
event horizon. It is convenient to use an orthogonal system of coordinates
associated with the falling observer. It can be readily verified that the
vectors,

$$
{\bf{e}}_{0}=\left\{ 1,-w\right\} ,{ \ \bf{e}}_{1}=\left\{
0,1\right\} , 
\eqno{(10)}
$$
\smallskip form an orthonormal basis; the first one being timelike. 

Notice that in this
basis, the vector \textbf{u} in Eq. (9), for $k = 0,$ and the Killing 
vector, \textbf{kv}, can be written,

$$
{\bf{u}}=\frac{\left( w,1\right) }{\left( w^{2}-1\right) ^{\frac{1}{%
2}}},{ \ \bf{kv}=\textbf{e}}_{0}+w{\textbf{e}}_{1}=\left(
1,w\right) 
\eqno{(11)}
$$
where we have adopted the following convention: $\{$v$_{0},$v$_{1}\}$, and $%
( $v$_{0},$v$_{1})$, denote vectors in the coordinate and orthonormal basis
respectively. It is clear that in Eq. (11), \textbf{u}$^{2}=-1$, and that 
\textbf{u}$\cdot $\textbf{kv}$=-k=0$. 

Returning to the case $|k|=1,$ the equations that need to be satisfied are,

$$
{\bf{u}}\cdot {\bf{kv}}=-k,{ }-\left( u^{t}\right)
^{2}+\left( u^{r}\right) ^{2}=-1,{ \ }|k|{ }=1.
\eqno{(12)}
$$

The four solutions to Eqs. (12) are,

$$
{\bf{u}}=-k\left( \frac{\left( w^{2}+1\right) }{\left(
w^{2}-1\right) },\frac{2w}{\left( w^{2}-1\right) }\right) ,{ \ }|k|=1 
\eqno{(13.a)}
$$

$$
{\bf{u}}=\left( 1,0\right) ,{ \ }k=1;{ \; }{ \bf{u}}%
=\left( -1,0\right) ,{ \ }k=-1 
\eqno{(13.b)}
$$
\smallskip 

We do not expect the particle's proper time to decrease while the coordinate
time increases, we are therefore left with only three physically meaningful
solutions. In the coordinate basis the components of these three solutions
are,

$$
{\bf{u}}=\left\{ 1,-w\right\} ,{ \bf{u}}=\left\{ -\frac{%
\left( w^{2}+1\right) }{\left( w^{2}-1\right) },+w\right\} ,{ \ }k=1. 
\eqno{(14.a)}
$$

$$
{\bf{u}}=\left\{ \frac{\left( w^{2}+1\right) }{\left(
w^{2}-1\right) },-w\right\} ,{ \ }k=-1.
\eqno{(14.b)}
$$
\smallskip

The solution with $k=1$, in Eq. (13.b), shows that \textbf{u} is the
velocity of a particle that is stationary with respect to the falling
observer, as expected from our choice of basic vectors in Eq. (10). The same
conclusion can be reached in the coordinate basis (cf. the first equation in
(14.a)), because it follows from Eq. (1.a), that $dr/d\tau =-w$.

Equation (14.b) represents a sinking particle with $e=1$ and has no physical
meaning above the event horizon because there, $dt/d\tau <0$. Conversely,
the second equation in (14.a) represents an outgoing particle with no
physical meaning below the event horizon. Here, the minimum descent velocity
is equal to $\left( w^{2}-1\right) ^{\frac{1}{2}}$ from Eq. (9), and $w$ in
Eq. (14.b) must satisfy the inequality $w>\left( w^{2}-1\right) ^{\frac{1}{2}%
}$, which is indeed the case.

\smallskip

\textbf{3. Particle Pair Creation and Hawking Effect}

\smallskip

The Killing vector, \textbf{kv} $=(1,0)$, is timelike above the event
horizon (out), and spacelike below (in). In the quantum creation of a pair
of particles, energy conservation requires that (Hartle, page 291),

$$
-{\bf{u}}_{in} \cdot {\bf{kv}}-{\bf{u}}_{out}\cdot 
{\bf{kv}}= k_{in}+k_{out}=0.
\eqno{(15)}
$$
\smallskip Here, \textbf{u}$_{in}$, \textbf{u}$_{out}$ are the
four-velocities of the created pair. Above the event horizon, $k_{out}$,
must be positive because it is proportional to the particle's energy
measured by an observer with velocity \textbf{kv}. Therefore $k_{in}=
-{\bf{u}}_{in} \cdot {\bf{kv}}$ must be negative and equal to 
$-k_{out}$. For values of $|k|=1$, it follows from Eq. (14.b),
that the velocity of the created particle below the event horizon must be, 
\textbf{u}$_{in}$=$\left\{ \left( w_{in}^{2}+1\right) /\left(
w_{in}^{2}-1\right) ,-w_{in}\right\} .$ From the time dependence of the
solutions in Eq. (14.a), it is apparent that \textbf{u}$_{out}$ must
then be taken equal to $\left\{ -\left( w_{out}^{2}+1\right) /\left(
w_{out}^{2}-1\right) ,{ }w_{out}\right\} ,$ which is the outgoing
solution for $k=1$. The particle above the event horizon cannot fall back
into the black hole. Because $w_{in}$ and $w_{out}$ are both very
approximately equal to one, it is straightforward to show that $%
u_{in}^{t}=u_{out}^{t}$ requires that Eq. (16.1) below, be satisfied.
Equation (16.2) follows from the relation, $e=(k^{2}+1)/2$,

$$
1-r_{in}=r_{out}-1,{ \ }e_{in}=e_{out},
\eqno{(16)}
$$
where $r_{in}$ and $r_{out}$ are the radial coordinates of the particles
forming the pair, an intuitively attractive result.

In Eq. (16.2), $e_{in}$ and $e_{out}$\ are the energies of the respective
particles divided by their mass. Because $e_{in}=e_{out}$, global energy 
conservation requires that the mass of the particle below the EH be equal 
to minus the mass of the particle above the EH, i.e., the energy of the 
particle below the event horizon must be negative in agreement with Schutz 
and Carlip's interpretation of the Hawking effect. The particle with negative
mass survives for a finite amount of time before reaching the center of the
black hole. But it is an unobservable particle \textit{and provides the
formalism with the necessary degrees of freedom that allows for the correct
interpretation of an observed particle at infinity escaping from the black
hole}. It would of course be of great interest to understand what happens
for values of $k$ such that $0 < k < 1$ because then the outside particle
cannot escape to infinity.

We calculate now the ratio $dt/d\tau $, where $t$ is the Schwarzschild time, 
and $\tau $\ is the proper time of the particles at the event horizon. 
From Eq. (3b) it follows that $dt/dt_{\left(metric\right) }$=%
$1/\left( 1-w^{2}\right) ,$ and then from Eq. (14a) we obtain for an outside
particle,

$$
\frac{dt}{d\tau }=\frac{\left( 1+w_{out}^{2}\right) }{\left(
1-w_{out}^{2}\right) ^{2}}=\frac{1}{2\left( 1-w_{out}\right) ^{2}}. 
\eqno{(17)}
$$

In a theory of particle creation, the proper time should play the relevant
role. Assume then that at the event horizon, $N$ particle-pairs are produced
in a time $d\tau $. Equation (17) shows that for the outside observer, $N$
particles will have been produced in the incomparably larger time, $dt=d\tau
/2\left( 1-w_{out}\right) ^{2},$ which suggests a weak observed productions
of particles. However, the Hawking radiation from a black hole is also very
weak. Field theory calculations show that black holes emits as though it
were a black body with temperature,

$$
k_{b}T=\frac{\hbar c^{3}}{\left( 8\pi GM\right) }, 
\eqno{(18)}
$$
the notation being standard. The temperature, $T$, is truly the physical
temperature of the black hole, not merely a quantity paying a role
mathematically analogous to temperature in the laws of black hole mechanics
(cf. Wald, p.12). Mini black holes excepted, the emission is weak, and the
creation of particles with finite mass, even neutrinos, must be weaker
still. However, as the temperature increases, during the final stages of
evaporation, the creation of particle-pairs with finite mass, could
conceivably, become important. It is clear however that the answer to this
issue \ lies far beyond the scope of this paper, and can only be obtained
with the help of field theories capable of calculating particle creation in
a curved space time (see, e.g., Parker).

\smallskip

\textbf{4. Acknowledgements}

\smallskip

Enlightening comments by Professor James Hartle on the first version of this
article are acknowledged. I am grateful to Drs. Bertrand Chavineau, Roger
Clark, and Robert Low for helpful discussions.

\smallskip

\smallskip \textbf{5. References}

\smallskip

Carlip, S.: Re: Hawking radiation and vacuum fluctuation,
http://sci.tech-archive.net/sci.physcsresearch/2004-10

\smallskip

Hartle, J.B.: \textit{Gravity, }Addison Wesley. California, (2003)

\smallskip

Parker, L.: Quantized Fields and Particle Creation in expanding Universes, 
\textit{Phys.Rev.} 183, 1057 (1969)

\smallskip

Schutz, B.: \ \textit{A First Course in General Relativity,} Cambridge
University Press (2007)\smallskip

\smallskip

Wald, R.M.: The Thermodynamics of Black Holes, \textit{Living Rev. Relativity%
} 4\textit{,} (2001), 6. URL (cited on 1,24,2008)
http://www.livingreviews.org/lrr-2001-6

\smallskip
\end{document}